\documentstyle[12pt]{article}

\begin{document}

\title{The shock waves in relativistic superfluid}
\author{G.V. Vlasov \thanks{%
E-mail: vs@itp.ac.ru}}
\maketitle

\begin{abstract}
We consider the discontinuities in a two-constituent relativistic
superfluid. In the acoustic limit they degenerate into the first and second
sound which are independent up to the second-order linear approximation.
Inclusion of the quadratic deviations relates to the small-amplitude shock.
Particularly we consider a plane shock at low temperature when the phonon
excitations contribute to the normal constituent. So we found the
generalization of the temperature increment and acoustic wave velocity in
relativistic superfluid. The fourth sound speed is also calculated.
\end{abstract}

\sloppy


\section{INTRODUCTION}

The first approach to the relativistic superfluid mechanics proposed by
Israel~\cite{Israel81} and Dixon~\cite{Dixon82} concerns with perfect
fluids. The method is useful for particular calculations, and as a general
model can be applied to relativistic superfluidity. For, strictly speaking,
the coupled constituents are not perfect fluids: any coupling results to
deviation from ideality. Or: the absence of coupling on microscopic level
implies thermo-isolation of the constituents~\cite{Vls97}. Nevertheless, the
attempt of taking into account the deviation from perfect fluid is not
senseless. The further development of Khalatnikov and Lebedev~ \cite{KL82}
includes the interaction between the superfluid and normal constituent. This
principle developed independently by Carter~\cite{Carter89} was gained in
recent works~\cite{CK92,CK94}. Although the main attention was directed to
the general formalism, including the equations of motion, rather than
applied problems, Carter and Langlois~\cite{CL95} have recently derived the
first and second sound speed in a superfluid with a phonon equation of state
of the normal constituent. This inspires us to discuss the shock wave
propagation in a two-constituent relativistic superfluid. While the
relativistic shock waves first considered by Taub~\cite{Taub48} have been
under detailed discussion~\cite{Taub78,Israel89,Cissoko92,Cissoko97} and the
shock waves in superfluid helium~ \cite{Khalatnikov89} were also
investigated, the present task bears a qualitatively new feature: we work
not in the frames of perfect fluid hydrodynamics, for it is impossible to
split the conserving particle number current into a conserving
''superfluid'' and ''normal'' parts~\cite{CL95} as it is in the Newtonian
limit~\cite{Khalatnikov89}.

We use the natural system of units ($\hbar =c_{light}=1$) and the metric
corresponding to the Minkowsky space with a metric tensor $diag(-+++)$.

\section{THE CONSERVATION LAWS}

The equations of relativistic superfluid mechanics are determined by the
Lagrangian $L$, whose infinitesimal variation is given by formula~\cite
{CK92,CK94,CL95} 
\begin{equation}
\delta L=\Theta _\varrho \/\delta s^\varrho \,-n^\varrho \/\delta \mu
_\varrho  \label{lagrangian}
\end{equation}
where the particle number vector $n^\varrho $ conjugated to the momentum
co-vector $\mu _\varrho $ obeys the conservation law 
\begin{equation}
\nabla _\varrho \,n^\varrho =0  \label{particles}
\end{equation}
and the entropy vector $s^\varrho $ conjugated to the thermal momentum
1-form $\Theta _\varrho $ is also conserved 
\begin{equation}
\nabla _\varrho \,s^\varrho =0  \label{entropy}
\end{equation}
till a shock wave appears. The energy-momentum tensor corresponding to the
Lagrangian $L$ has the form 
\begin{equation}
T_\nu ^\varrho =n^\varrho \mu _\nu +s^\varrho \Theta _\nu +\Psi g_\nu
^\varrho  \label{tensor}
\end{equation}
with the pressure function 
\begin{equation}
\Psi =L-\Theta _\varrho \/s^\varrho \,\/  \label{pressure}
\end{equation}
The conservation law of the energy-momentum tensor~(\ref{tensor}) 
\begin{equation}
\nabla _\varrho \,T_\nu ^\varrho =0\,\,\/  \label{tens-cons}
\end{equation}
leads to the equation of motion of the normal constituent 
\begin{equation}
s^\varrho \left( \nabla _\varrho \Theta _\nu -\nabla _\nu \,\Theta _\varrho
\right) =0\,\,\/  \label{motion}
\end{equation}
and the irrotationality condition 
\begin{equation}
\nabla _\varrho \,\mu _\nu -\nabla _\nu \,\mu _\varrho =0\,\,\/
\label{irrot}
\end{equation}
Vectors $n^\varrho $ and $\mu ^\varrho $(also $s^\varrho $ and $\Theta
^\varrho $) are not colinear. For the particle number vector $n^\varrho $
determines the Eckart rest frame, while $\mu ^\varrho $ determines similarly
the superfluid rest frame and that does not coincide with the former. The
Lagrangian $L\left( \mu ,y,s\right) $ depends on three invariants, them
being 
\begin{equation}
s^2=-s^\varrho s_\varrho \,\,\/  \label{s}
\end{equation}
for the normal rest frame entropy density; for the cross product given by 
\begin{equation}
y^2=-s^\varrho \mu _\varrho \,\,\/  \label{y}
\end{equation}
and for the effective mass variable (i.e. the chemical potential in the
superfluid rest frame) 
\begin{equation}
\mu ^2=-\mu ^\varrho \mu _\varrho  \label{mu}
\end{equation}
The secondary variables $n^\varrho $ and $\Theta ^\varrho $ can be expressed
through the primary variables $\mu ^\varrho $ and $s^\varrho $ according to
the formula 
\begin{equation}
\left( 
\begin{array}{c}
n^\varrho \\ 
\Theta ^\varrho
\end{array}
\right) =\left( 
\begin{array}{cc}
B & -A \\ 
A & C
\end{array}
\right) \left( 
\begin{array}{c}
\mu ^\varrho \\ 
s^\varrho
\end{array}
\right)  \label{expr}
\end{equation}
with the coefficients obtained immediately by differentiation of the
Lagrangian 
\begin{equation}
B=2\frac{\partial L}{\partial \mu ^2}\qquad C=-2\frac{\partial L}{\partial
s^2}\qquad A=-\frac{\partial L}{\partial y^2}  \label{coeff}
\end{equation}
On the other hand we can write

\begin{equation}  \label{2-expr}
\left( 
\begin{array}{c}
n^\varrho \\ 
s^\varrho
\end{array}
\right) =\left( 
\begin{array}{cc}
F & Q \\ 
Q & G
\end{array}
\right) \left( 
\begin{array}{c}
\mu ^\varrho \\ 
\Theta ^\varrho
\end{array}
\right) =\left( 
\begin{array}{cc}
B+A^2/C & -A/C \\ 
-A/C & 1/C
\end{array}
\right) \left( 
\begin{array}{c}
\mu ^\varrho \\ 
\Theta ^\varrho
\end{array}
\right)
\end{equation}
and calculate new coefficients through the pressure function \cite{Vls97} 
\begin{equation}  \label{coeff-new}
F=2\frac{\partial \Psi }{\partial \mu ^2}\qquad G=2\frac{\partial \Psi }{%
\partial \Theta ^2}\qquad Q=\frac{\partial \Psi }{\partial z^2}
\end{equation}

However, a more convenient way is to obtain all parameters~(\ref{coeff}) in
terms of (\ref{s}), (\ref{y}) and (\ref{mu}) and express them in terms of
the relative translation speed between the superfluid and normal reference
frames 
\begin{equation}  \label{speed}
w^2=1-\frac{\mu ^2s^2}{y^4}
\end{equation}
and the effective temperature

\begin{equation}  \label{theta}
\Theta ^2=-\Theta ^\varrho \Theta _\varrho
\end{equation}
instead of invariants $y$ and $s$. At low temperature the phonon-like
excitations with energetic spectrum $\omega =cp$, where 
\begin{equation}  \label{snd}
c^2=\frac n\mu \frac{d\mu }{dn}
\end{equation}
and the latter is the sound speed contribute to thermodynamical functions of
the normal constituent. The ''phonon'' Lagrangian has the form~\cite{CL95} 
\begin{equation}  \label{lagr-pres}
L=P-3\psi \qquad \psi =\frac k{c^{1/3}}\left[ s^2+\left( c^2-1\right) \frac{%
y^4}{\mu ^2}\right] ^{2/3}=\frac{\pi ^2}{90c^3}\frac{\Theta ^4}{\left[
1-w^2/c^2\right] ^2}\/
\end{equation}
where the pressure of excitations is

\begin{equation}  \label{psi}
\psi =\frac{s_s\Theta }4
\end{equation}
and 
\begin{equation}  \label{ss}
s_s=\frac s{\sqrt{1-w^2}}=\frac{\pi ^2}{15c^3}\frac{\Theta ^4}{\left[
1-w^2/c^2\right] ^2}
\end{equation}
is the entropy $s_s$ in the superfluid reference frame, while $k$ is a
definite constant. The coefficients (\ref{coeff}) then are

\begin{equation}  \label{low-cf}
A=\frac{1-c^2}\mu \frac \Theta {c^2-w^2}\qquad C=-\frac 1{s_s}\frac \Theta
{c^2-w^2}
\end{equation}
For an ultra-relativistic spectrum of excitations ($c\rightarrow 1$) the
term $A$ disappears.

\section{THE DISCONTINUITIES}

The discontinuities in superfluid helium have been first discovered by
Khalatnikov \cite{Khalatnikov89}. Discussing the discontinuities in a
relativistic superfluid we shall follow the standard formalism of the
relativistic shock waves~\cite{Taub78}. However, since we deal with a
strongly self-interacting medium and two-constituent at that, a standard
perfect fluid theory is impossible. So we have to think of a new extended
formalism, combining the theory of relativistic shock waves~\cite
{Taub78,Israel89,Cissoko92,Cissoko97} and the relativistic superfluid
mechanics \cite{CK92,CK94,CL95}. In the Newtonian limit this method must be
reduced to the non-relativistic theory of discontinuities in helium \cite
{Khalatnikov89}, and in the acoustic limit this method must give the first
and the second sound in a relativistic superfluid~\cite{CL95}. The amplitude
of discontinuities is assumed to be not very large, since superfluidity is
expected to take place on both sides of the front with no phase transition.

Let the hypersurface $\Sigma $ be the front of the discontinuity and vector $%
\lambda ^\varrho $ be a unit (space-like) normal to it. The conservation
laws (\ref{particles}), (\ref{tens-cons}) entail the conditions 
\[
\left[ \lambda _\varrho \/n^\varrho \right] =0\qquad \qquad \left[ \lambda
_\varrho \/T_\nu ^\varrho \right] =0\/ 
\]
i.e. 
\begin{equation}  \label{d-particles}
n^{\perp }=\lambda _\varrho \/n_{+}^\varrho =\lambda _\varrho
\/n_{-}^\varrho \/\/
\end{equation}
\begin{equation}  \label{d-tensor}
\lambda _\varrho \/T_{+\/\nu }^{\/\/\varrho }=\lambda _\varrho \/T_{-\/\nu
}^{\/\/\varrho }\/
\end{equation}
where indexes + and - relate to the quantities ahead of and behind the
discontinuity, respectively, and the square brackets imply the change across
the front of a discontinuity. Eq. (\ref{d-particles}) conveys the continuity
of the orthogonal part (marked always by index $\perp $) of the particle
number current.

We cannot make use of the entropy conservation law (\ref{entropy}) since it
does not take place in the shock waves. For that reason the results of Ref.%
\cite{Vls97} pertain for a most general two-constituent system (but not a
superfluid itself), while the discussion of small perturbations suffers no
restrictions concerning eq.~(\ref{entropy}). Indeed, the results for shock
waves in a two-constituent system and in a superfluid will coincide when the
magnitude of the relevant discontinuities tends to zero.

For a two-constituent superfluid the additional equation is the
irrotationality condition~(\ref{irrot}) which yields 
\[
\lambda _\varrho \,\mu _{+\nu }-\lambda _\nu \,\mu _{+\varrho }=\lambda
_\varrho \,\mu _{-\nu }-\lambda _\nu \,\mu _{-\varrho }\/ 
\]
Multiplying it by $\lambda ^\varrho $ and, then, by a unit vector $\eta
^\varrho $ orthogonal to $\lambda ^\varrho $, we get an important relation 
\begin{equation}  \label{d-irrot}
\mu _{+}^{\parallel }=\mu _{-}^{\parallel }
\end{equation}
(where index $\parallel $ denotes the tangential part of any quantity,
particularly $\mu ^{\parallel }=\eta ^\varrho \mu _\varrho $). This is the
very condition for conservation of irrotational motion passing a plane shock
wave~\cite{VK97}. It is clear that for a multi-constituent system the
irrotational motion conserves at both sides of the shock wave if condition~(%
\ref{irrot}) takes place for each constituent.

Substituting the expression (\ref{tensor}) and (\ref{d-particles}) in (\ref
{d-tensor}), we obtain 
\begin{equation}
n^{\perp }\left( \mu _{+\nu }-\mu _{-\nu }\right) +s_{+}^{\perp }\Theta
_{+\nu }-s_{-}^{\perp }\Theta _{-\nu }=-\left( \Psi _{+}-\Psi _{-}\right)
\lambda _\nu  \label{2-tensor}
\end{equation}
Multiplying this equation by $\lambda ^\nu $, we get 
\begin{equation}
n^{\perp }\left( \mu _{+}^{\perp }-\mu _{-}^{\perp }\right) +s_{+}^{\perp
}\Theta _{+}^{\perp }-s_{-}^{\perp }\Theta _{-}^{\perp }=-\left( \Psi
_{+}-\Psi _{-}\right)  \label{d-pressure}
\end{equation}
Then, multiplying eq.(\ref{2-tensor}) by the unit vector $\eta ^\nu $, in
light of (\ref{d-irrot}), we get 
\begin{equation}
n^{\perp }\left( \mu _{+}^{\parallel }-\mu _{-}^{\parallel }\right)
+s_{+}^{\perp }\Theta _{+}^{\parallel }-s_{-}^{\perp }\Theta _{-}^{\parallel
}=0\/  \label{dicont}
\end{equation}
This important relation determines the types of discontinuities. Then, the
irrotationality condition (\ref{d-irrot}) than coincides with the condition
of strong discontinuity in the superfluid constituent which occurs when $%
n^{\perp }\neq 0$. As for the discontinuity in the normal component, in
light of (\ref{d-irrot}), it is determined merely by the single relation 
\begin{equation}
s_{+}^{\perp }\Theta _{+}^{\parallel }=s_{-}^{\perp }\Theta _{-}^{\parallel
}\/  \label{d-normal}
\end{equation}
Thus the constraint (\ref{d-irrot}) corresponds to ordinary shock waves,
while (\ref{d-normal}) beseems to a ''temperature'' discontinuity of the
second sound type.

As a particular instance of another type of discontinuity we consider a
vortex sheet in superfluid~\cite{Parts-etal94}. Since the vortex sheet
separates the whole space into domains where the superfluidity takes place,
the irrotationality condition~(\ref{irrot}) does not hold in the global
sense and we cannot establish the constraint (\ref{d-irrot}) at both sides
of the sheet. Therefore, the tangential discontinuities are possible. The
conservation law~(\ref{d-particles}) then yields $n_{+}^{\perp
}=n_{-}^{\perp }=0$ that determining weak or slip-stream discontinuity: the
particle number flow across the front of discontinuity equals zero, indeed,
no matter crosses the hypersurface of the discontinuity, i.e. this
hypersurface is made up of streamlines of the fluid.

\section{A PLANE SHOCK IN FLAT SPACE}

Let the discontinuity propagates along the axis $x^1$. We choose the unit
normal $\lambda ^\nu =(0,1,0,0)$ - and the medium at rest before the front.
As a rule one used to practice with the rest frame co-moving the front, so
that the fluid flows in the front with the velocity which is equal to that
of a shock wave. Hereby, the relevant vectors and co-vectors may be
presented as 
\begin{equation}  \label{cold}
n^\varrho =n\left( \/\!\sqrt{1+\varphi ^2}\/,\/\varphi \/,\/0,\/0\right)
\/\qquad \qquad \mu _\varrho =\mu \left( -\/\sqrt{1+\xi ^2}\/\/,\xi
,\/0,\/0\right) \/
\end{equation}
\begin{equation}  \label{normal}
s^\varrho =s\left( \/\!\sqrt{1+\alpha ^2}\/\/,\/\alpha \/,\/0,\/0\right)
\/\qquad \qquad \Theta _\varrho =\Theta \left( -\/\sqrt{1+\beta ^2}%
\/\/,\/\beta \/,\/0,\/0\right) \/
\end{equation}
Since the medium ahead of front is at rest, the relative velocity $w_{-}$
equals zero, $y_{-}^2=\mu _{-}s_{-}$, while behind the shock 
\[
y_{+}^2=\frac{\mu _{+}s_{+}}{\sqrt{1-w^2}}\/\/ 
\]
Also 
\begin{equation}  \label{ab}
\alpha _{-}=\beta _{-}=\varphi _{-}=\xi _{-}\equiv x\/\/
\end{equation}
Hence, the velocity of the shock is determined as 
\begin{equation}  \label{u}
u=\frac x{\sqrt{1+x^2}}\/\/
\end{equation}
Substituting our definitions (\ref{cold}), (\ref{normal}) and (\ref{ab}) in
eqs. (\ref{d-particles}), (\ref{d-irrot}), (\ref{d-normal}) and (\ref
{d-pressure}) we, firstly obtain

\begin{equation}  \label{u-particles}
\/n_{+}\varphi =\/n_{-}x
\end{equation}
The rest equations, in view of (\ref{u-particles}), will be 
\begin{equation}  \label{u-irrot}
\mu _{+}\sqrt{1+\xi ^2}=\mu _{-}\sqrt{1+x^2}
\end{equation}
\begin{equation}  \label{u-normal}
\alpha s_{+}\Theta _{+}\sqrt{1+\beta ^2}=xs_{-}\Theta _{-}\/\sqrt{1+x^2}
\end{equation}
\begin{equation}  \label{u-pressure}
n_{-}x\left( \xi \mu _{+}-x\mu _{-}\right) +\alpha \beta s_{+}\Theta
_{+}-x^2s_{-}\Theta _{-}=-\Psi _{+}+\Psi _{-}
\end{equation}
The parameter $\varphi $ incorporates only in eq. (\ref{u-particles}) and it
can be calculated as soon as the rest unknowns are found. Thus, in eqs. (\ref
{u-irrot}), (\ref{u-normal}) and (\ref{u-pressure}) the unknowns are: the
four parameters $\alpha $, $\beta $, $\xi $, $x$ and three invariants (\ref
{s}), (\ref{y}), (\ref{mu}) behind the shock on which the pressure $\Psi
_{+} $ depends. The pressure behind the shock can be expressed through $\mu
_{+}$, $\Theta _{+}$, and through the relative velocity for which we use the
notation $w$. The formula

\begin{equation}  \label{w}
\frac 1{\sqrt{1-w^2}}=\sqrt{1+\alpha ^2}\sqrt{1+\xi ^2}-\alpha \xi
\end{equation}
relates the later quantity with $\alpha $ and $\xi $. Our goal is to find
the velocity of the shock wave $u$ for a single parameter given behind the
shock. Without the loss of generality $\mu _{+}$ can be chosen for this
parameter. Thus, there are six unknowns in four equations (\ref{u-irrot}), (%
\ref{u-normal}), (\ref{u-irrot}), (\ref{w}). The rest two relations follow
from (\ref{expr}) or (\ref{2-expr})

\begin{equation}  \label{u-n}
\/\/n_{-}x=F_{+}\mu _{+}\xi +Q_{+}\Theta _{+}\beta
\end{equation}

\begin{equation}  \label{u-s}
s_{+}\alpha =Q_{+}\mu _{+}\xi +G_{+}\Theta _{+}\beta
\end{equation}
with the coefficients (\ref{coeff}) calculated for the state behind the
front.

The knowledge of the equation of state in explicit form is necessary for
calculation of the right-hand side of eq. (\ref{u-pressure}) and the
coefficients in eqs. (\ref{u-n}) and (\ref{u-s}).

\section{A LOW TEMPERATURE CASE}

The low-temperature equation of state was derived by Carter and Langlois 
\cite{CL95}. In view of (\ref{psi}), (\ref{ss}), (\ref{low-cf}) the
expressions (\ref{u-normal}), (\ref{u-pressure}), (\ref{u-n}), (\ref{u-s})
take the form 
\begin{equation}  \label{l-normal}
\bar \psi \sqrt{1-w^2}\alpha \sqrt{1+\beta ^2}=x\,\sqrt{1+x^2}
\end{equation}
\begin{equation}  \label{l-pressure}
x\left( \xi \bar \mu -x\right) +\tau \left( 4\alpha \beta \bar \psi \sqrt{%
1-w^2}-4x^2+\bar \psi -1\right) =-\frac 1\Gamma \left( \bar P-1\right)
\end{equation}
\begin{equation}  \label{l-n}
\bar \mu \bar F\xi +\left( 1-c_{+}^2\right) \tau \frac{\bar \psi }{\bar \mu }%
\beta = x
\end{equation}
\begin{equation}  \label{l-s}
\sqrt{1-w^2}\alpha =\left( 1-c_{+}^2\right) \xi +\left( c_{+}^2-w^2\right)
\beta
\end{equation}
where 
\[
\tau =\frac{\psi _{-}}{\mu _{-}n_{-}}\qquad \qquad \Gamma =\frac{\mu
_{-}n_{-}}{P_{-}}\qquad \qquad \bar F=F_{+}/\left( \frac{n_{-}}{\mu _{-}}%
\right)
\]

At low temperature we have the estimations \cite{CL95,Vls97}

\begin{equation}  \label{estim}
F=\frac n\mu +O\left( \Theta ^4\right) \qquad G\sim \Theta ^2\qquad Q\sim
\Theta ^3
\end{equation}
implying that $\tau =O\left( \Theta ^4\right)$ and, hence, equations (\ref
{u-pressure}) and (\ref{u-n}) approximately (up to the terms $O\left(
\Theta^4\right)$) coincide with their zero-temperature version.

\section{THE SOUND, STRONG AND SMALL-AMPLITUDE SHOCK WAVES}

If all parameters behind the front tend to their values ahead, the shock
becomes a sound wave. Since the entropy in the sound wave is conserved, we
can apply formalism \cite{Vls97} achieved for a two-constituent relativistic
medium with the conserved particle currents of both constituents. In the
linear approximation both methods lead to the same result, namely from the
system (\ref{u-irrot}), (\ref{u-normal}), (\ref{u-pressure}), (\ref{w}), (%
\ref{u-n}), (\ref{u-s}) we obtain an equation for two branches of sound at
arbitrary temperature which is analogous to that derived by Carter \cite
{Carter89} and, under assumption $A^2=o\left( C\right) $, splits into 
\begin{equation}
u_I^2=\frac{-B}{B+\mu B_\mu +sA_\mu }  \label{first}
\end{equation}
\begin{equation}
u_{II}^2=1+\frac{sB_s+\mu A_s}C  \label{second}
\end{equation}
and reduces, in the low temperature limit, to the first and the second sound
speed, respectively \cite{CL95}: $u_I=c$, $u_{II}=c\, \sqrt{3}$. Here for
any variable $V$ we used the notation 
\begin{equation}
V_s=\frac{\partial V}{\partial s}+\frac \mu {2y}\frac{\partial V}{\partial y}%
\qquad V_\mu =\frac{\partial V}{\partial \mu }+\frac s{2y}\frac{\partial V}{%
\partial y}  \label{diff}
\end{equation}
However, if we omit $A$, the second sound speed calculated by formula (\ref
{second}) with the phonon Lagrangian (\ref{lagr-pres}) of Carter and
Langlois \cite{CL95} will be $u_{II}=1/3$ instead of obvious $u_{II}=c/3$.
Because the Lagrangian (\ref{lagr-pres}) is derived for the two-fluid theory
with non-zero cross term; while the Lagrangian of thermal excitations of the
Israel theory \cite{Israel89} differs from (\ref{lagr-pres}); although the
relative translation speed $w$ between the constituents is presented in both
approaches. So each Lagrangian is useful in the theory to which it does
belong.

In order to find the velocity of a small-amplitude shock we rewrite eqs.(\ref
{u-irrot}), (\ref{l-normal}), (\ref{l-pressure}), (\ref{w}), (\ref{l-n}), (%
\ref{l-s}) in the second-order approximation. After tedious calculations we
find the velocity increment 
\begin{equation}
\frac{\Delta u_{II}}{u_{II}}=\frac{1-c^2}{1+c^2}\frac{\Delta \Theta }\Theta
\label{uuu}
\end{equation}
of the shock corresponding to the second sound when the temperature
increment $\Delta \Theta$ tends to zero. In turn, the latter gives rise to a
finite relative speed $w$ behind the shock, since 
\begin{equation}
\frac{\Delta \Theta }\Theta =\frac w{c\sqrt{3}\sqrt{1-c^2/3}}
\label{t-2order}
\end{equation}
In the non-relativistic limit the equations (\ref{uuu}) and (\ref{t-2order})
yield well known expressions \cite{Khalatnikov89}. The shock occurs ahead of
the second sound, as is in superfluid helium at low temperature.

So, in the acoustic limit the solution splits into two independent branches $%
u_I$ and $u_{II}$ corresponding to the first and the second sound. The first
branch describes wave propagation through the medium which behaves as a
perfect fluid composed of two constituents whose pressure and enthalpy are $%
P+\psi$ and $\mu n+4\psi$ respectively. While the constituents in the waves
of the second branch move independently, a counterflow appears: $w\neq0$. In
general, a ''mixed'' solution occurs, and the temperature increases together
with the chemical potential.

The estimations (\ref{estim}) imply that the first sound and the relevant
shock wave coincide roughly with usual discontinuity in the cold
constituent.A great pressure jump $\Delta \Psi$ is produced inevitably by
the change in pressure of the superfluid constituent $P$, since the
contribution of the normal constituent $\psi$ is small. Hence, in view of
eqs. (\ref{l-pressure}) and (\ref{l-n}) we conclude that strong shock waves
at low temperature propagate with the speed $u=u_0+O\left( \Theta ^4\right)$
which approximately equals to the speed of a usual shock wave in cold
constituent $u_0$, but always $u_I>u_{II}$. A more precise result is

\begin{equation}  \label{u1}
u_I^2=u_0^2\left\{ 1-\tau u_0^2\left( 4-\Gamma \right) \right\}
\end{equation}
\begin{equation}  \label{u2}
u_{II}^2=u_0^2\left\{ 1-\tau u_0^2\left( 4-\Gamma \sqrt{1-w^2}\right)
\right\}
\end{equation}

The second-sound discontinuities should be regarded as ''moderate'' for
intermediate values of $u_{II}$. This takes place if relative changes in the
superfluid and normal variables are of the same order and they can be of the
same order if they do not access $\tau$ greatly. For a superfluid matter of
neutron stars and phonon equation of state it is easy to estimate $\tau
\sim10^{-9}$. For this particular example we performed calculations with an
ultra-relativistic superfluid matter. The sound speed in this medium equals
exactly to $c=1/\sqrt{3}$, and approximately it is the first sound, while
the speed of the second sound $c_{II}=c/\sqrt{3}$. For a not very small
pressure change the velocity of the shock wave will be merely $u_0$. The
dependence of the relative translation speed $w$ on $u_0$ is given in table
1.

\begin{table}[tbp]
\tablename{~1.~The gamma-factor $\gamma _w=w/\sqrt{1-w^2}$ of the relative
translation speed vs the Mach number $M=u/c$ of the shock wave and the
pressure change} \label{void}\vspace{.5cm}
\par
\begin{tabular}{|r|r|r|r|r|r|}
\hline
$\Psi_+/\Psi_--1$ & $1.1$ & $1.5$ & $2$ & $5$ & $10$ \\ \hline
$M$ & $1.024$ & $1.10$ & $1.17$ & $1.35$ & $1.46$ \\ \hline
$\gamma_w$ & $6.7\cdot 10^{-3}$ & $.031$ & $.065$ & $.256$ & $.476$ \\ \hline
\end{tabular}
\end{table}

Although the second sound velocity attains to the saturation value $u_0$,
the relative translation speed $w$ grows with the growth of the shock wave
intensity.

\section{THE FOURTH SOUND}

The fourth sound takes place when the normal constituent is restrained by
some external agent, while the sound propagates through the superfluid
constituent. We cannot use the equation of motion (\ref{motion}), but the
conservation laws (\ref{particles}), (\ref{entropy}) and the irrotationality
condition (\ref{irrot}) will determine the fourth sound speed. If the sound
wave propagates in the direction determined by vector $\lambda ^\nu
=\left(u,1,0,0\right)$ the change of gradient of arbitrary quantity $V$ is
proportional to its infinitesimal change $\hat V$ \cite{Carter89}: $%
\left[\nabla _\nu \hat V\right] =\hat V\,\lambda _\nu $. Thereby, we write - 
\begin{equation}
-u\hat n^0\,+\hat n^1=0  \label{inf-n}
\end{equation}
\begin{equation}
-u\hat \mu _1\,-\hat \mu _0=0\,\,  \label{inf-mu}
\end{equation}
\begin{equation}
-u\hat s^0\,+\hat s^1=0  \label{inf-s}
\end{equation}
instead of (\ref{particles}), (\ref{entropy}), (\ref{irrot}) and (\ref
{motion}). In the reference frame co-moving with the normal constituent we
put 
\begin{equation}
s^\varrho =s\left( 1,0,0,0\right) \qquad \mu _\varrho =\mu _n\left(
-1,w,0,0\right)  \label{smu}
\end{equation}
where $\mu _n=\mu /\sqrt{1-w^2}$ is the chemical potential in the ''normal''
reference frame. Since the discontinuities propagate through the superfluid
constituent, there must be 
\begin{equation}
\hat s^\varrho =0  \label{s0}
\end{equation}
Equations (\ref{inf-n}), (\ref{inf-mu}), (\ref{s0}) analogous to the
relevant non-relativistic set \cite{Khalatnikov89} determine the speed of
the fourth sound. Requiring the vanishing determinant of the system (\ref
{inf-n}), (\ref{inf-mu}) we get a quadratic equation for $u$. While the
speed of the first and the second sound is determined by a 4-order system 
\cite{CL95}, the fourth sound speed follows from two equations. At zero
temperature the speed of the first and the fourth sound are obtained by the
same equations (\ref{inf-n}) and (\ref{inf-mu}) and coincide exactly with
the sound speed $c$ in the superfluid constituent. The difference appears at
finite temperature on account of the relationship \cite{CL95} 
\[
\hat n^\varrho =B^{\varrho \nu }\hat \mu _\nu +C^{\varrho \nu }\hat s_\nu 
\]
between the infinitesimal discontinuities in (\ref{inf-n}), (\ref{inf-mu})
and the temperature dependence of matrices in (\ref{relat}). For the phonon
equation of state we get the explicit formula

\begin{equation}
u_{IV}=\/c+c\,\frac{\/\rho _n}{\rho _s}\left( -\frac 13+\frac{17}6c^2-3c^4-%
\frac{c\,\mu }{1+c^2}\frac{\partial c}{\partial \mu }+\frac 23c\mu ^2\frac{%
\partial ^2c}{\partial \mu ^2}\right)  \label{relat}
\end{equation}
which generalizes the relevant non-relativistic relation \cite{Khalatnikov89}%
, where $\rho _n$ and $\rho _s$ is the normal and the superfluid
energy-density respectively \cite{CL95}.

\section{CONCLUSION}

Summarizing the results obtained in the present study, we emphasize the
formulae (\ref{d-particles}), (\ref{d-irrot}), (\ref{d-normal}), (\ref
{d-pressure}) which determine the propagation of discontinuities through a
two-constituent relativistic superfluid in the general case. For a plane
shock wave seven equations (\ref{u-particles}), (\ref{u-irrot}), (\ref
{u-normal}), (\ref{u-pressure}), (\ref{w}), (\ref{u-n}) and (\ref{u-s}) with
seven unknowns must be solved. In the acoustic limit these equations reduce
to formulae (\ref{first}), (\ref{second}) for the first and the second
sound. At low temperature the system (\ref{u-particles}), (\ref{u-irrot}), (%
\ref{u-normal}), (\ref{u-pressure}), (\ref{w}), (\ref{u-n}) and (\ref{u-s})
reduces to (\ref{u-particles}), (\ref{u-irrot}), (\ref{l-normal}), (\ref
{l-pressure}), (\ref{w}), (\ref{l-n}) and (\ref{l-s}). The velocity of
strong shock waves is given by (\ref{u1}) and (\ref{u2}), while (\ref{uuu})
and (\ref{t-2order}) describ the change of parameters in a weak shock wave.
As for perspectives and applications, the shock waves and spin-isospin sound
in the nuclear matter are worth to be discussed in future.

\end{document}